\documentclass{elsart}

\usepackage{graphicx}
\usepackage{epstopdf}

\usepackage{amssymb}
\usepackage{longtable}

\begin{document}
\def\pbp{\rm{pbp}}
\def\pp{\rm{pp}}
\begin{frontmatter}
\title{Precise measurement of the absolute fluorescence yield of the 337 nm band \\in atmospheric gases}



\author[usc]{{\bf AIRFLY Collaboration}: M. Ave},
\author[CZ1]{M. Bohacova},
\author[Chi]{E. Curry},
\author[AqU]{P. Di Carlo},
\author[RomeU]{C. Di Giulio},
\author[Chi]{P. Facal San Luis},
\author[IKall]{D. Gonzales},
\author[FNAL]{C. Hojvat},
\author[IMAPP]{J. H\"orandel},
\author[CZ2]{M. Hrabovsky},
\author[AqU]{M. Iarlori},
\author[IKall]{B. Keilhauer},
\author[IKall]{H. Klages},
\author[IKall]{M. Kleifges},
\author[FNAL]{F. Kuehn},
\author[Chi]{S. Li},
\author[Chi]{M. Monasor},
\author[CZ1]{L. Nozka},
\author[CZ1]{M. Palatka},
\author[AqU]{S. Petrera},
\author[Chi]{P. Privitera\corauthref{cor1}},
\ead{priviter@kicp.uchicago.edu}
\author[CZ1]{J. Ridky},
\author[AqU]{V. Rizi},
\author[Chi]{B. Rouille D'Orfeuil},
\author[Ors,AqU]{F. Salamida},
\author[CZ1]{P. Schovanek},
\author[IKall]{R. Smida},
\author[ANL]{H. Spinka},
\author[Mun]{A. Ulrich},
\author[RomeU]{V. Verzi},
\author[Chi]{C. Williams}
\footnotesize
\address[usc]{Universidad de Santiago de Compostela, Departamento de F\'isica de Part\'iculas, \\ Campus Sur, Universidad, E-15782 Santiago de Compostela, Spain}
\address[CZ1]{Institute of Physics of the Academy of
    Sciences of the Czech Republic,\\ Na Slovance 2, CZ-182 21 Praha 8,
    Czech Republic}
\address[Chi]{University of Chicago, Enrico Fermi Institute \& Kavli Institute for
    Cosmological Physics, 5640 S. Ellis Ave., Chicago, IL 60637, USA}
\address[AqU]{Dipartimento di Scienze Fisiche e Chimiche dell'Universit\`{a}
    de l'Aquila and INFN, Via Vetoio, I-67010 Coppito, Aquila, Italy}
\address[RomeU]{Dipartimento di Fisica
    dell'Universit\`{a} di Roma Tor Vergata and Sezione INFN,\\ Via
    della Ricerca Scientifica, I-00133 Roma, Italy}
\address[IKall]{Karlsruhe Institute of Technology, Postfach 3640, D - 76021 Karlsruhe, Germany}
\address[FNAL]{Fermi National Accelerator Laboratory, Batavia, IL 60510, USA}
\address[IMAPP]{IMAPP, Radboud University Nijmegen, 6500 GL Nijmegen, The Netherlands}
\address[CZ2]{Palacky University, RCATM, Olomuc, Czech Republic}
\address[Ors]{Institut de Physique NuclŽaire d'Orsay (IPNO), UniversitŽ Paris 11, CNRS-IN2P3, Orsay, France}
\address[ANL]{Argonne National Laboratory, Argonne, IL 60439 United States}
\address[Mun]{Physik Department E12, Technische Universit\"{a}t M\"{u}nchen,
James Franck Str. 1, D-85748 Garching, Germany}
\corauth[cor1]{corresponding author}
\newpage
\begin{abstract}
A measurement of the absolute fluorescence yield of the 337 nm nitrogen band, relevant to ultra-high energy cosmic ray (UHECR) detectors, is reported. Two independent calibrations of the fluorescence emission induced by a 120 GeV proton beam were employed: Cherenkov light from the beam particle and calibrated light from a nitrogen laser. The fluorescence yield in air at a pressure of 1013~hPa and temperature of 293~K was found to be $Y_{337}  = 5.61\pm 0.06_{stat} \pm 0.21_{syst}  ~\rm{photons/MeV}$. 
When compared to the fluorescence yield currently used by UHECR experiments, 
this measurement improves the uncertainty by a factor of three, and has a significant impact on the determination of the energy scale of the cosmic ray spectrum. 
\end{abstract}

\end{frontmatter}

\section{Introduction}
\label{sec:intro}
   A well established technique for the detection of Ultra-High Energy ($\gtrsim 10^{18}$~eV) Cosmic Rays (UHECRs) - first successfully employed by  the Fly's Eye~\cite{flyseye} and HiRes~\cite{hires} experiments -  is based on nitrogen fluorescence light emission induced by Extensive Air Showers (EAS). Excitation of atmospheric nitrogen by EAS charged particles results in fluorescence emission, mostly in the wavelength range between 300 and 430~nm. This UV light is measured as a function of time and incoming direction by photomultiplier cameras at the focus of large (few m$^2$) mirrors. Fluorescence telescopes measure the longitudinal EAS development in the atmosphere, which provides unique information on the primary cosmic ray's type and a calorimetric measurement of its energy.  

The fluorescence light yield along the EAS depends on the air pressure, temperature and humidity at the point of emission. In addition, wavelength-dependent atmospheric attenuation  affects the light intensity reaching the telescope. Thus, the intensities of the fluorescence bands must be known for atmospheric conditions corresponding to the EAS development in the atmosphere, which ranges between about 2 km and 15 km above sea level. Early measurements of the fluorescence yield include those with low-energy stopped-particles in air by Bunner~\cite{bunner} and with electrons in air by Davidson and O'Neil~\cite{david}. A number of other experiments have made measurements of the fluorescence light yield pertinent to EAS \cite{kakim}  \cite{nagano1} \cite{belz} \cite{colin} \cite{gorod} \cite{airlight}. 

The AIRFLY (AIR FLuorescence Yield) Collaboration has carried out an extensive program of measurements to significantly improve the precision on the fluorescence light yield. The fluorescence emission was studied as a function of the kinetic energy, ranging from keV to GeV, of particle beams at several accelerators \cite{airflyedep}. The relative intensities of 34 fluorescence bands in the wavelength range from 284  to 429~nm, together with their pressure dependence, were reported in \cite{airflyp}. The temperature and humidity dependence of the main fluorescence bands was also measured~\cite{airflyth}. These measurements have provided the most complete and consistent set of fluorescence yield data for UHECR calibration, establishing the {\it relative} spectral line dependence on atmospheric parameters.  In this paper, a precise measurement of the {\it absolute} fluorescence yield - namely the number of photons emitted per MeV of energy deposited - is reported. This result will impact the current generation of UHECRs experiments, the Pierre Auger Observatory~\cite{auger} and the Telescope Array~\cite{telaray}, as well as a future large observatory or the proposed space-based JEM-EUSO~\cite{jemeuso}, which rely on fluorescence detection for their absolute energy scale. In fact, the fluorescence yield is presently a major contribution to the systematic uncertainty of the cosmic ray energy (14\% for the Pierre Auger Observatory~\cite{augersyst} and 11\% for the Telescope Array~\cite{tasyst}, with both experiments quoting a total systematic uncertainty of 22\%). 

In a standard approach to a photon yield measurement, the absolute quantum efficiency of the photon detector ultimately limits the systematic uncertainty. AIRFLY follows a different approach, first presented in~\cite{airflyabsolute}, 
based on the comparison of the fluorescence yield to Cherenkov light emitted in the same experimental setup. 
  Measurements are performed with a narrow band optical filter centered around the main nitrogen fluorescence line at 337 nm. Since the fluorescence yield is measured relative to the known Cherenkov yield, the absolute value of the photon detection efficiency is not essential, with a considerable reduction of the systematic uncertainties. 
In this work, several improvements have been implemented with respect to the preliminary test in~\cite{airflyabsolute}, most notably the use of a single particle beam at the Test Beam Facility of the Fermi National Accelerator Laboratory (FNAL) and of an integrating sphere for optimal light collection. 
Measurements were mostly performed with 120 GeV~protons\footnote{Notice that, at these energies, the nature of the primary particle does not affect the fluorescence emission of the 337~nm band, which is induced by the secondary electrons produced in the gas~\cite{arquerosyield}. In fact, the cross section for electron impact excitation of the $\rm{C^3\Pi_u}$ state of the nitrogen molecule (responsible for the 337~nm emission)  becomes completely negligible for energies above 10~keV, and direct excitation by protons is also negligible~\cite{protonexc}.}.
In addition, a measurement of the fluorescence yield was performed  which employed a nitrogen laser as a calibrated light source. When the two methods are combined, a total uncertainty of 4\% is achieved. A preliminary account of this measurement was presented in \cite{airflybejing}. 

This paper is organized as follows.  Section~\ref{sec:expmethod} presents the apparatus and experimental methods. Simulations are described in Section~\ref{sec:geant4sim}.  Data analysis is presented in Section~\ref{sec:eventsel}. The procedure for the fluorescence yield measurement is described in Section~\ref{sec:abscherana}. Results of the fluorescence yield measurement, including systematic uncertainties, with the Cherenkov calibration and with the laser calibration are given in Sections~\ref{sec:abscher}~and~\ref{sec:abslaser}, respectively. The combined measurement is presented in Section~\ref{sec:combined}, and conclusions are drawn in Section~\ref{sec:conclusions}.

\section{Apparatus and experimental methods}
\label{sec:expmethod}
In this Section, the AIRFLY apparatus is presented. Details of the integrating sphere, a fundamental component of the  experiment, are given. The procedures for the measurement of fluorescence and Cherenkov light are illustrated. The laser setup, which provided a calibration light source for the fluorescence measurement, is also described. 
Lastly, the measurement of the 337 nm fluorescence line yield in nitrogen to that in air is presented. 
 
\subsection{The AIRFLY apparatus at the FNAL Test Beam Facility}
\label{sec:mtest}
\begin{figure*}[th]
\centering \includegraphics[width=3.7in,angle=-90]{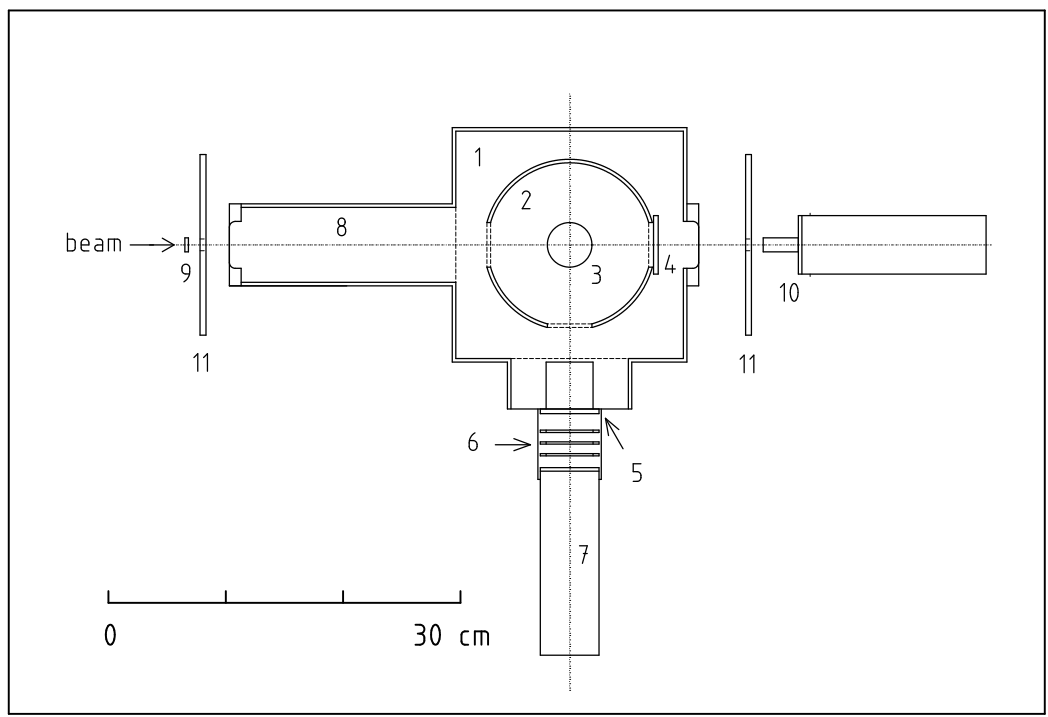}
  \caption{The AIRFLY experimental setup: 1) stainless steel chamber; 2) integrating sphere; 3) top port; 4) plug of exit port; 5) 337 nm interference filter; 6) acceptance apertures; 7) PMT; 8) extension for beam particle path; 9) scintillator disk counter; 10) Cherenkov counter, with acrylic rod and photomultiplier; 11) veto counters. }
\label{fig:layout}
\end{figure*}

The measurements were performed at the FNAL Test Beam Facility\footnote{http://www-ppd.fnal.gov/FTBF/}, which provided a  primary 120~GeV proton beam from the Main Injector and secondary beams of muons, pions, electrons and positrons of lower energy. The proton beam was selected in the majority of the measurements for its larger multiplicity, smaller beam profile and negligible contamination. 
Every minute, the beam is spilled in the experimental area during 4~s, with particles grouped in trains of bunches. Trains are separated in time by 10~$\mu$s, and bunches within each train are separated by 19~ns. Typical conditions for data taking with the proton beam were 30~bunches per train with a multiplicity of $2 \cdot 10^5$ particles per spill. The beam profile, monitored by wire chambers placed before and after the AIRFLY apparatus, was typically 3~mm x 4~mm (r.m.s.). 
The layout of the apparatus is shown in Fig.~\ref{fig:layout}. 
A scintillator disk counter (10~mm diameter and 3~mm thickness) tagged the beam at the entrance of the AIRFLY chamber. A Cherenkov counter - a UV-transparent acrylic cylindrical rod of 10~mm diameter and 30~mm long - tagged the beam at the fluorescence chamber's exit. 
Cherenkov light produced in this counter by the beam particles provided a signal of few ns time resolution and very good single particle resolution. Two large 150~mm x 150~mm~scintillator counters, with a central hole of 10~mm diameter for the beam, vetoed off-axis particles. The fluorescence chamber was placed between the two veto pads. 
The gas chamber was a 3~mm thick stainless steel cube of 200~mm side length.  Flanges were provided for beam windows, shutters, gauges, gas inlet and vacuum. The beam entrance window, made of 0.1~mm thick aluminum, was installed on an upstream extension of the chamber to provide an additional length of 18~cm for Cherenkov light production. The exit window, also of 0.1~mm thick aluminum, was mounted on an extension to the downstream side of the chamber. An integrating sphere located at the center of the gas chamber (Section \ref{sec:sphere}) collected the light produced by the beam in the chamber. Light out of the integrated sphere passed through a 337~nm interference filter (bk Interferenzoptik, model bk-337 1-10-C2, 70\% peak transmission) 
that served at the same time as a chamber window. A mechanical optical shutter was incorporated with the filter. The shutter could be closed and opened remotely to alternate background measurements during data taking. 
A photomultiplier tube (Hamamatsu, model H7195P), hereafter called PMT, with good single photoelectron resolution was used for photon detection. The optical field of view was defined by a 40~mm diameter acceptance cylinder placed between the integrating sphere's port and the filter, and by circular apertures of the same size placed in front of the PMT photocathode. The chamber was internally covered with black UV-absorbing material to avoid stray light.
An additional scintillator counter vetoing off-axis particles that could hit the PMT and a second H7195P photomultiplier with
a closed shutter for background monitoring, both placed on the side of the PMT, are not shown in Fig.~\ref{fig:layout}.
The chamber and the counters were mounted on an optical breadboard for precise alignment. The breadboard was placed on a remotely controlled position table, which allowed the alignment of the apparatus with the beam by maximizing the rates of the entrance scintillator disk and exiting Cherenkov counters. 
A laser system (Section \ref{sec:lasersystem}) was routinely inserted in the apparatus for calibration purposes. 
A remotely controlled gas handling and vacuum system was installed. The chamber pressure, temperature and humidity were measured. Typical pressure and temperature ranges during the fluorescence measurements were $1013 \pm 10$~hPa and $293 \pm 2$~K. The same high purity dry gases, 99.9999\% pure nitrogen and an air-like mixture (79\% nitrogen, 21\% oxygen), of our previous measurements~\cite{airflyp}~\cite{airflyth} were used. Pure helium was used for background measurements.

\begin{figure*}[th]
\centering \includegraphics[width=5.8in]{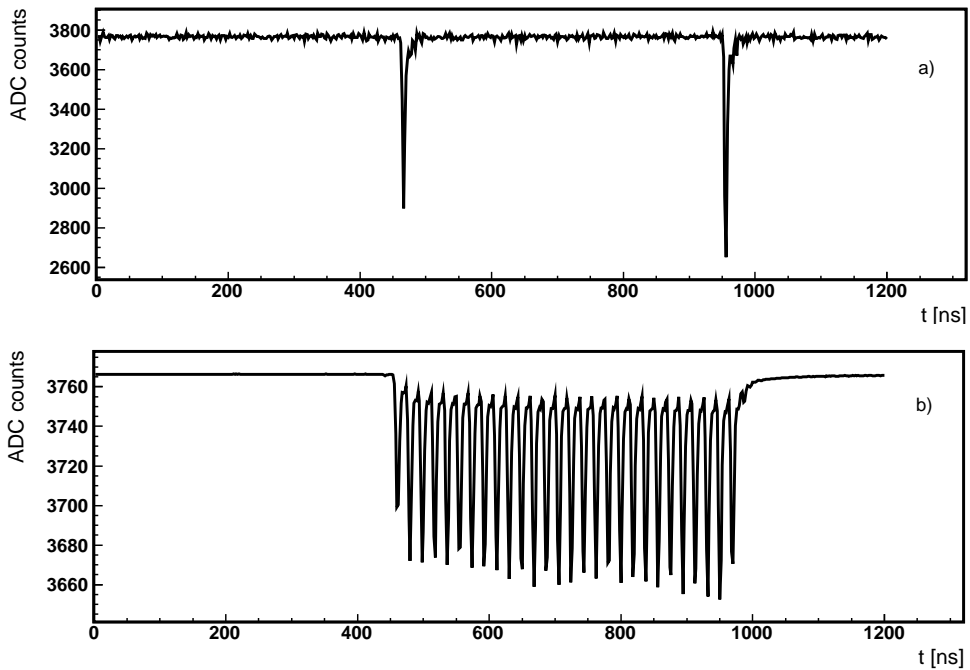}
  \caption{Digitized ADC trace of the Cherenkov counter: a) trace recorded for one event trigger, with two fast pulses; b) ADC trace averaged over all event triggers in one spill, showing the bunch structure of the beam.  }
\label{fig:cherevent}
\end{figure*}

The data acquisition was based on a VME system. An event trigger was formed from the coincidence of a train trigger gate, issued in time with the arrival of each train of bunches, and a single particle trigger gate from a beam monitoring scintillator counter. The Test Beam Facility provided both signals. A single detected particle within a train caused the signals from the apparatus scintillator counters and PMTs to be digitized by a 12-bit 500~MHz Flash ADC (model SIS3350, Struck Innovative Systeme). 
For each event trigger, 600~samples equivalent to 1.2~$\mu\rm{s}$ - thus containing the entire train of bunches - were saved in the FADC memory. The train trigger gate was used as time reference for the trigger signal, and was found to be stable at the level of 1~ns. As an example, the ADC trace of the Cherenkov counter for one event trigger is shown in Fig. \ref{fig:cherevent}(a), presenting two fast pulses produced by particles in two separate bunches. The ADC trace averaged over all event triggers in one spill is shown in Fig. \ref{fig:cherevent}(b), with the time structure of the bunches clearly distinguishable.

The module's internal memory of 128~MSamples/channel was used to store the data during the 4~s spill without dead time. Data for up to 1~GByte (the exact number depending on the number of triggers during the spill) were read out and saved to disk in about 40~s between spills. While demanding in terms of data transfer and storage, this data acquisition strategy provided accurate information for each triggered train, allowing flexibility, redundancy and detailed systematic studies in the subsequent offline analysis. Data taking was organized in runs integrating 30~spills for about 30~min duration, to guarantee stable conditions.  A fast analysis of a sizable fraction of the data was performed online, providing several monitoring variables like rates, time distributions and ADC integral distributions. A total of 7~TByte of data was collected during two dedicated test beam periods. 

\subsection{Integrating sphere and data taking configurations}
\label{sec:sphere}
The integrating sphere - a hollow sphere with an internal surface highly diffusive in the UV - was a fundamental component of the AIRFLY apparatus. The sphere collected the light over almost the $4\pi$ solid angle. 
Since only about 0.2 fluorescence photons of the 337~nm band were produced from a single beam particle in the chamber,  light collection over a large solid angle maximized the signal over background and guaranteed a reasonable data taking time. Also, the light is made isotropic and unpolarized after several diffusion bounces on the sphere's surface, producing at the detection port a Lambertian light output which is practically independent of the original light distribution. Thus, the isotropic fluorescence light and the highly directional Cherenkov or laser light present an identical distribution at the entrance of the 337~nm filter, greatly reducing systematic uncertainties.  

The integrating sphere was formed by  two hollow aluminum half-spheres of 15.2~cm diameter and 3~mm thickness. Four ports of 38.1~mm diameter were machined on the sphere: two of them along the beam direction (beam entrance and exit), one at the side, orthogonal to the beam direction (photon detection), and the last port at the sphere's top. The sphere's internal surface was covered with a 0.5~mm thick GORE$\texttrademark$ DRP$^{\textregistered}$ \cite{gore} reflector. This proprietary material, made of a micro porous structure of PTFE, has a very high diffusive reflectance ($\approx 99.5\%$) around 337~nm, providing a sphere throughput $\approx 40\%$ higher than that of a commercial sphere made of granular PTFE. The thin GORE$\texttrademark$ DRP$^{\textregistered}$  reflector, when compared to the many mm thick PTFE walls of a commercial sphere, minimized Cherenkov light produced by particles passing through the diffusive material.
Two remotely operated shutters could close the downstream exit and the top ports of the sphere, while the entrance and the side ports were always open. The port plugs on the shutters were made of the same GORE$\texttrademark$ DRP$^{\textregistered}$  reflector, hereafter called diffuser for the sake of simplicity. 

Two different configurations of the chamber were used to detect either fluorescence or Cherenkov light. 
In the {\it Fluorescence configuration}, the exit port of the integrating sphere was open, while the top port was closed. In this way, the highly directional Cherenkov light, emitted along the particle path in the chamber, does not hit the sphere's diffusive surface. After exiting the sphere through the downstream port, the light is absorbed by the black inner coating of the chamber, acting as a Cherenkov light dump. Thus, only fluorescence light emitted along the particle path is collected by the integrating sphere and measured by the PMT viewing the sphere's side port. 
In the {\it Cherenkov configuration}, the top port was open, while the exit port was closed. With this arrangement, the Cherenkov light is diffused back into the sphere by the exit port plug, the integrating sphere collecting then both the Cherenkov and the fluorescence light.
Notice that, by opening or closing the top port, the total reflecting surface of the integrating sphere is kept the same in both configurations - always three open ports - allowing for an accurate comparison of the fluorescence and Cherenkov measurements.

Measurements were also performed with  aluminized mylar mirrors replacing the diffuser plugs at the top and exit ports.
The mirror reflective surface was placed at an angle of $160^\circ$ with respect to the beam direction, thus reflecting the Cherenkov light back into the sphere. The use of a thin mirror rather than a diffuser plug eliminated a major background - Cherenkov photons produced by beam particles in the PTFE of the exit port plug. The top port was also fitted with a mirror, to keep the sphere port's symmetry between the Fluorescence and Cherenkov configurations.

\subsection{UV Laser calibration}
\label{sec:lasersystem}
\begin{figure*}[t]
\centering \includegraphics[width=3.25in,angle=-90]{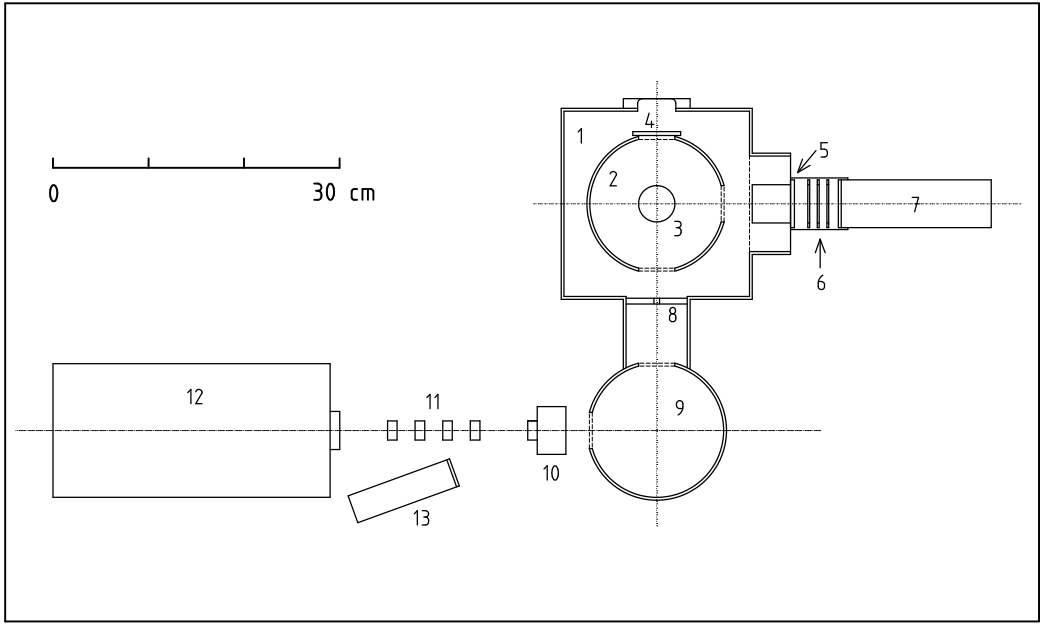}
  \caption{The laser calibration system integrated in the AIRFLY setup: 8) and 10) collimators;  9) laser calibration sphere; 11) optical attenuators; 12) 337~nm nitrogen laser; 13) $\rm{PMT}_{\rm{laser}}$, to monitor laser intensity. For all other components, see caption of Fig. \ref{fig:layout}.}
\label{fig:laser}
\end{figure*}

The laser calibration system consisted of  a 337~nm nitrogen laser (Spectra-Physics, model VSL-337ND-S), producing light pulses of $\approx3$~ns width at up to 20~Hz repetition rate,  and a PTFE integrating sphere (SphereOptics, model SPH-6Z-4) referred to as {\it laser calibration sphere} in the following. A sketch of the laser system during a calibration run at the Test Beam Facility is shown in Fig. \ref{fig:laser}. 
The laser light was attenuated by a set of filters and collimators before entering the laser calibration sphere. A photomultiplier tube, $\rm{PMT}_{\rm{laser}}$,  which detected light reflected by an attenuation filter  at $\approx 180^\circ$ from the laser beam, was used to provide a fast trigger signal of the laser pulse.  Light exited the laser calibration sphere  through a 6~mm diameter hole at the end of a 15.5~cm long collimator tube mounted on the sphere's exit port. The efficiency of the laser calibration sphere was determined from the ratio of the light pulse energy  measured at the exit of the collimator tube  to that measured at the entrance of the sphere. A silicon energy probe (LaserProbe inc., model RjP-465) with fJ sensitivity  was used to measure the laser pulse energy. The efficiency of the laser calibration sphere was measured to be $(2.49\pm0.02) \cdot 10^{-6} $, where the 0.8\% uncertainty takes into account the probe linearity (0.5\%) and the reproducibility of the result (0.6\%). 
In a laser calibration run, the system was mounted at the front of the AIRFLY chamber, with the collimator tube facing the entrance port of the integrating sphere (Fig. \ref{fig:laser}). The laser pulse energy at the entrance of the laser calibration sphere was $\approx 5$~pJ, measured with 5\% uncertainty by the  RjP-465 probe, which gave a calibrated beam of $\approx 20$~photons per pulse entering the AIRFLY sphere. For each laser pulse, the PMT signal was digitized by the same electronics chain of the fluorescence measurements (Section \ref{sec:mtest}) and a single photoelectron signal was searched for (Section \ref{sec:eventsel}). A calibration constant, $S_{\rm{laser}}$, was then obtained  by counting the number of detected photoelectrons per 337 nm photons entering the AIRFLY sphere (shortened as pp in the following). A typical measurement yielded  $S_{\rm{laser}} \approx 11\cdot 10^{-4}~\pp $.   
The laser beam energy  was measured by the  silicon probe at the beginning and at the end of each 15~min calibration run. The stability of the laser pulse energy during the calibration measurement was monitored by recording the signal from $\rm{PMT}_{\rm{laser}}$, and found to be typically better than 1\%. Several laser calibration runs were taken each day at the end of the fluorescence measurements.   

\subsection{Nitrogen to air fluorescence ratio}
\label{sec:n2air}
With the beam multiplicity available at the Test Beam Facility,  the detected rate of fluorescence photons in air was very small. 
For a precise measurement, long data taking runs would have been necessary, which were unpractical in the context of a test beam. 
Thus, most of the measurements were performed with pure nitrogen, where the fluorescence yield in the 337~nm band is almost ten times larger.  
The air fluorescence yield was then obtained by using the independently measured 
ratio of the 337~nm fluorescence in pure nitrogen to that in air, $r_{\rm{N_2}}$.

AIRFLY measured the nitrogen to air fluorescence ratio as a function of pressure in \cite{airflyp}, obtaining $ r_{\rm{N_2}} = 7.38 \pm 0.08$ at 1000 hPa. The ratio could be slightly different in the present setup, which included the integrating sphere and a new 337~nm interference filter. 
Thus, a dedicated measurement was performed with a 18~$\mu$Ci $^{241}$Am source. The radioactive source was placed at 3~cm distance from the entrance port of the sphere. Alpha particles from  $^{241}$Am have a range of about 4~cm in air, thus depositing most of their energy close to the entrance port. The air fluorescence spectrum induced by alpha particle excitation \cite{alphafluor} is consistent with the AIRFLY measurement with electrons \cite{airflyp}. Measurements of the PMT single photoelectron rates were then performed in pure nitrogen and air.
In order to keep the same path length of the alphas in the gas, and thus the same spatial distribution of the fluorescence emission, measurements in nitrogen and air were taken at slightly different pressures to compensate for the gas densities.  From these measurements, a ratio 
 \begin{equation}
r_{\rm{N_2}} = 7.45 \pm 0.07 
\label{eq:rn2}
\end{equation}
at 1000~hPa is derived, in very good agreement with our previous measurement~\cite{airflyp}. 

\section{Monte Carlo simulations}
\label{sec:geant4sim}
A full simulation of the AIRFLY apparatus was performed in the Geant4~\cite{geant4} framework. In this Section, details of the photon detection components - the integrating sphere, the 337 nm filter and the PMT - are given. 
Also, the implementation of the fluorescence and Cherenkov yields in the simulation is described, and studies of the energy deposited by protons in gas are presented. 
Finally, results of the simulation relevant for the fluorescence yield measurement are given. 

\subsection{Photon detection efficiency}
\label{subsec:photondet}
The diffuser coating the inner surface and the port plugs of the integrating sphere was simulated as a 0.5~mm thick PTFE layer, diffusing incident light with reflectivity $\epsilon_{\rm{diff}}$ and a Lambertian angular distribution.  The value of $\epsilon_{\rm{diff}}$ was chosen to reproduce the sphere's throughput measured in a dedicated experiment. A 337~nm laser  beam was shot into the sphere, and the laser light power was measured at the sphere's entrance and at 11.5~cm distance from the side port. The ratio of these two measurements provided a benchmark efficiency of the sphere to be reproduced by the simulation. 
The efficiency,  $6.04 \cdot 10^{-4}$, measured with a 1\% uncertainty which includes the photodiode non-linearity and the reproducibility of the result, was reproduced in a full simulation of this experiment for $\epsilon_{\rm{diff}}=99.43 \%$. The sphere with aluminized mirror plugs was characterized by a similar set of measurements, which were reproduced by introducing in the  simulation a value of  $87.2 \%$ for the reflectivity of the aluminized mirror at 337~nm.

The wavelength dependence of the integrating sphere's throughput was also included in the simulation.
For this purpose, the reflectance of the diffuser was measured as a function of wavelength with a standard reflectance setup (integrating sphere AvaSphere-50 and spectrometer AvaSpec-2048).  
A weight was then implemented in the simulation for each diffusion process occurring inside the sphere. The aluminized mylar mirror reflectivity was also measured, and found to be constant within the bandwidth of the 337~nm interference filter. Since photons bounce on average about 20 times on the diffuser inside the sphere before exiting from the side port, the wavelength dependence of the sphere's efficiency is determined by the diffuser properties, and it did not change significantly with the mirror plugs. 

The transmittance of the 337~nm interference filter was measured as a function of wavelength and angle of incidence of the light with a Lambda EZ210 UV/Vis spectrophotometer. This dependence must be included in the simulation, given the broad spectrum of Cherenkov light and the strong dependence of interference filter's transmission on angle of incidence.
Measurements in steps of 0.1~nm and for 13 incident angles between $0^\circ$ and $20^\circ$ (the maximum angle of incidence in the AIRFLY geometry) were performed. Data were linearly interpolated to cover all wavelengths and angles in the simulation.    

The quantum efficiency of the PMT was provided by the manufacturer as a function of wavelength. The absolute detection efficiency of the PMT was implemented in the simulation as the product of the quantum efficiency by a collection efficiency of 85\% (a typical value for this type of photomultipliers according to the manufacturer). 
The absolute efficiency of the PMT as a function of wavelength was also independently measured with a dedicated setup, to be reported elsewhere \cite{PMTcalib}, yielding a result consistent with the 10\% systematic uncertainty quoted by the manufacturer.  

\subsection{Photon yields and energy deposit}
\label{subsec:edepsim}
In the Geant4 simulation, fluorescence photons were produced proportional to the energy deposited in a given air volume, and then emitted isotropically. The proportionality of the fluorescence yield to the energy deposit has been experimentally verified \cite{belz} \cite{colin} \cite{airflyedep} and is justified by the physics processes occurring at the microscopic level \cite{arquerosyield}. Photons were sampled from the air fluorescence spectrum measured by AIRFLY \cite{airflyp}, with a nominal yield for the 337~nm band of 5.15~photons/MeV of energy deposit in air at a pressure of 1013~hPa and temperature of 293~K. 

Particles, both primaries and  secondaries, are individually tracked in the simulation and energy is deposited along their path according to standard electromagnetic and hadronic processes. To speed up simulations, when the particle's range becomes shorter than one mm, tracking is stopped and the particle's energy is deposited locally.
This procedure provides a realistic simulation of the spatial distribution of the energy deposit (and consequently of the fluorescence emission), which is very important to avoid biases in the measurement of the fluorescence yield~\cite{arquerossyst}.

\begin{figure*}[th]
\centering \includegraphics[width=5.5in]{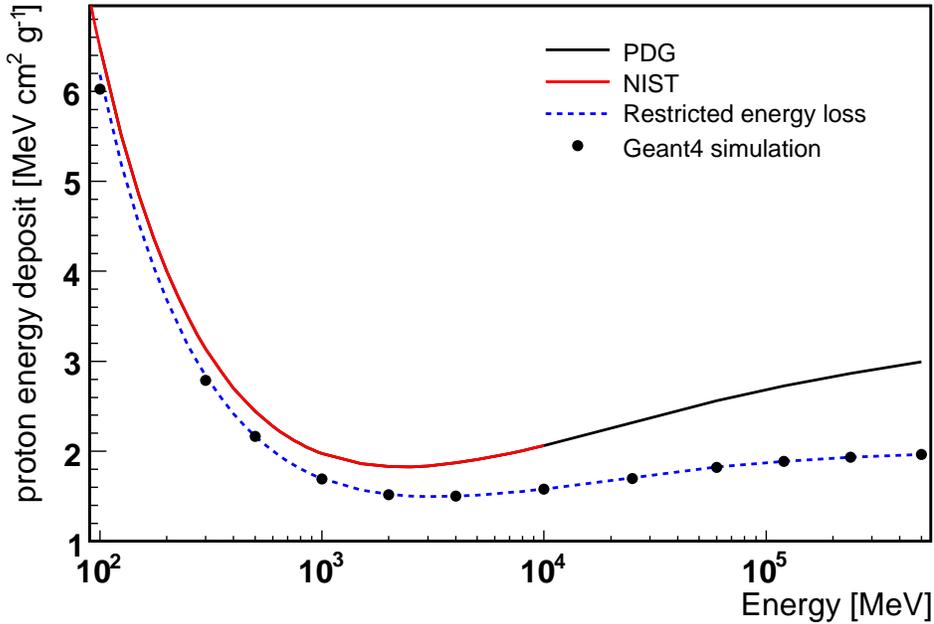}
  \caption{Simulation of proton energy deposit in nitrogen. Standard energy loss parameterizations from the Particle Data Group and NIST are compared with that implemented in Geant4 (full lines, differences undistinguishable on this scale).  Closed dots represent the energy deposited by a proton in the integrating sphere's volume according to Geant4; superimposed as a dashed line is the restricted proton energy loss calculated for a maximum energy transfer of 100~keV.}
\label{fig:simedep}
\end{figure*}

Dedicated simulations were performed to check the Geant4 model for the energy deposit. Results are shown in Fig.~\ref{fig:simedep}. Reference calculations of the proton energy loss in nitrogen from the Particle Data Group~\cite{pdgeloss} and NIST~\cite{pstar} are accurately reproduced by the simulation. Also, simulations of the energy deposit in the integrating sphere's nitrogen volume are well matched by the restricted proton energy loss~\cite{pdgeloss} calculated for a maximum energy transfer of 100~keV to secondary particles. This result is consistent with the path of $\delta$~rays being limited by the size of the sphere, since electrons of 100~keV kinetic energy have a range of 14~cm in nitrogen. 
Notice that a 120~GeV proton has negligible radiative losses, which makes the calculated energy loss directly comparable with the simulated energy deposit.  
The Geant4 simulation was checked against a different simulation~\cite{arquerosyield} where energy losses are implemented at the microscopic level according to excitation and ionization cross sections. In particular, the energy deposited by relativistic electrons in the integrating sphere's nitrogen volume was compared. A systematic uncertainty of 2\% on the simulation of the energy deposit was conservatively estimated, which includes the accuracy in the reference data of the Bethe-Bloch stopping power. 
This estimate is consistent with a detailed study of the simulation of the energy deposit presented in~\cite{arquerossyst1}.

The emission of Cherenkov photons is a standard electromagnetic process in the Geant4 simulation, with an absolute yield determined by the index of refraction of the gas. Parameterizations of the index of refraction as a function of wavelength for nitrogen~\cite{n_nitrogen1}~\cite{n_nitrogen2}, air~\cite{n_air} and helium~\cite{n_helium}  were implemented in the Geant4 simulation. Several checks were performed on the accuracy of these parameterizations. The nitrogen index of refraction taken from~\cite{n_nitrogen1} is based on measurements between 468~nm and 2060~nm, while that taken from~\cite{n_nitrogen2} is based on measurements between 145~nm and 270~nm. Both parameterizations fit data between 230~nm and 490~nm~\cite{n_helium} extremely well, differing only 0.2\% from the index of refraction measured at 334~nm. The corresponding Cherenkov yields, integrated over the bandwidth of the 337~nm filter, were found to differ by less than 0.1\%. For the refraction index of air, the uncertainty of Ciddor's parameterization~\cite{n_air} between 230~nm and 1700~nm is a few $10^{-4}$. From these studies, the systematic uncertainty on the absolute Cherenkov yield is estimated to be less than 0.1\%. 

\subsection{Simulation results}
\label{subsec:simresults}
\begin{figure*}[th]
\centering \includegraphics[width=5.5in]{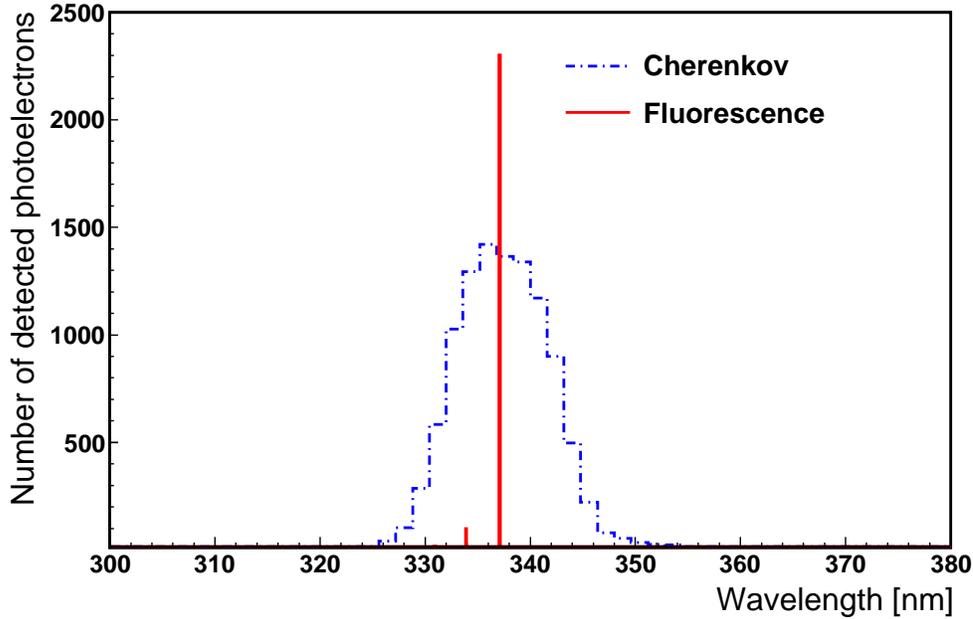}
  \caption {Simulation of the AIRFLY experiment: number of detected photoelectrons from fluorescence (full line) and Cherenkov (dot-dashed line) photons produced by $10^7$ protons in air.   }
\label{fig:effdet}
\end{figure*}

Monte Carlo simulations of the experiment relevant for the fluorescence measurement were performed. Protons of 120~GeV energy were generated at 20~m distance from the apparatus, and tracked through the air volume and a beam tagging scintillator of the Test Beam line before reaching the AIRFLY apparatus. The beam profile was simulated according to the measured spread. For each simulated event, the energy deposited by the proton in the AIRFLY counters was recorded. Fluorescence and Cherenkov photons emitted along the proton's path inside the chamber were tracked and diffused within the integrating sphere. Photons exiting from the side port of the sphere may pass through the 337~nm filter and reach the PMT, where a photoelectron was detected according to the PMT efficiency. 
An example of simulation is shown in Fig.~\ref{fig:effdet}, where the number of detected photoelectrons from fluorescence and Cherenkov photons produced by $10^7$ protons in air is given as a function of wavelength.  

The rates of single photoelectrons, expressed in units of photoelectrons per beam particle (\pbp), are given in Table \ref{tab:rateMC} as $S^{\rm{MC}}_{\rm{Fl}}$ and  $S^{\rm{MC}}_{\rm{Ch}} $ for the Fluorescence and Cherenkov simulated configurations, respectively. 
Results for both the diffuser plugs and the mirror plugs are reported. 
Also, the laser calibration constant, $S^{\rm{MC}}_{\rm{laser}} $ (Section~\ref{sec:lasersystem}), in units of photoelectrons per 337~nm photon (\pp), was derived from a Monte Carlo simulation of the laser calibration.
The statistical uncertainty on all simulation results is less than 1\%.  

\begin{table}[ht] 
 \centering \begin{tabular}{|c|c|c|c|} \hline 
       &   $S^{\rm{MC}}_{\rm{Fl}}$ ($10^{-4}~\pbp$)     &     $S^{\rm{MC}}_{\rm{Ch}} $ ($10^{-4}~\pbp$)   &   $S^{\rm{MC}}_{\rm{laser}} $ ($10^{-4}~\pp$)  \\ [0.5ex]	%
\hline 
diffuser plugs& 2.442 & 10.53 & 11.58 \\
 \hline 
mirror plugs  & 2.131 & 8.267 & \\
 \hline 
\end{tabular}
\caption{Summary of Monte Carlo simulation results.}
\label{tab:rateMC} 
 \end{table}

\section{Data analysis}
\label{sec:eventsel}
\begin{figure*}[th]
\centering \includegraphics[width=5.7in]{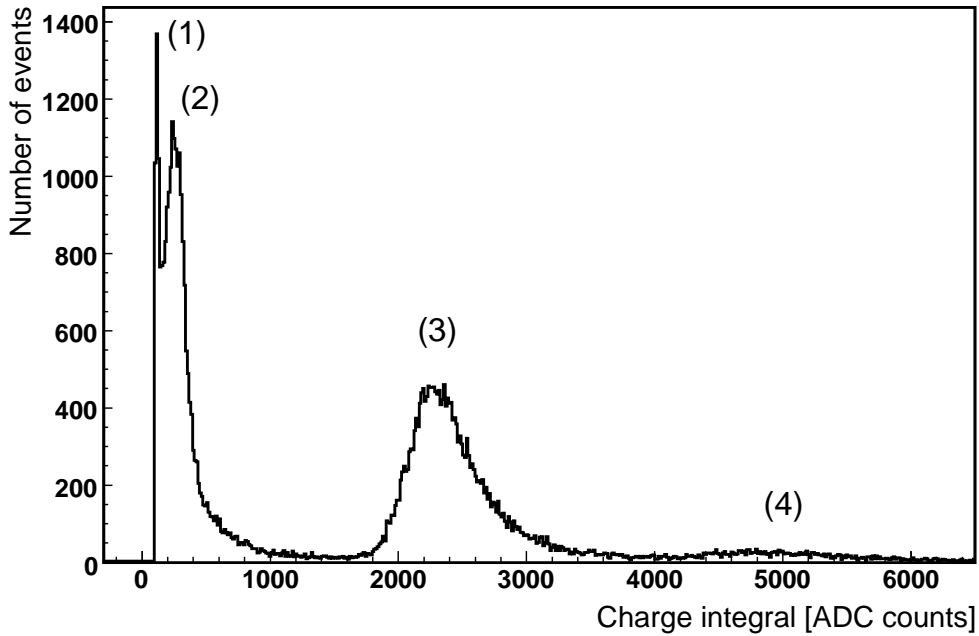}
  \caption{ Spectrum of the Cherenkov counter: (1) pedestal, truncated for better visualization; (2) signal corresponding to beam particles hitting the photomultiplier tube's window outside the 1~cm diameter acceptance of the rod; (3) signal corresponding to a single beam particle passing through the rod; (4) signal corresponding to two beam particles.}
\label{fig:cherrodspectr}
\end{figure*}

The rate of single photoelectrons at the PMT in time coincidence with a single beam particle was measured to determine the fluorescence yield. The selection of single beam particles was strongly based on the properties of the Cherenkov counter tagging the beam at the exit of the chamber.  For illustration, a spectrum of the Cherenkov counter, obtained by integrating the charge around the known positions of the bunches (Fig.~\ref{fig:cherevent}), is shown in  Fig.~\ref{fig:cherrodspectr} for a data run where the beam was purposely misaligned.
The peak~(3) around 2300~ADC counts, corresponding to a single beam particle passing through the rod, is well separated from the pedestal~(1) and from the signal for two beam particles~(4). Also, a peak~(2) is present corresponding to beam particles hitting the photomultiplier tube's window outside the 1~cm diameter acceptance of the rod. 
Thus, the Cherenkov counter provided a very clean single particle tagging, in both charge resolution and space.

\begin{figure*}[!t]
\centering \includegraphics[width=5.52in]{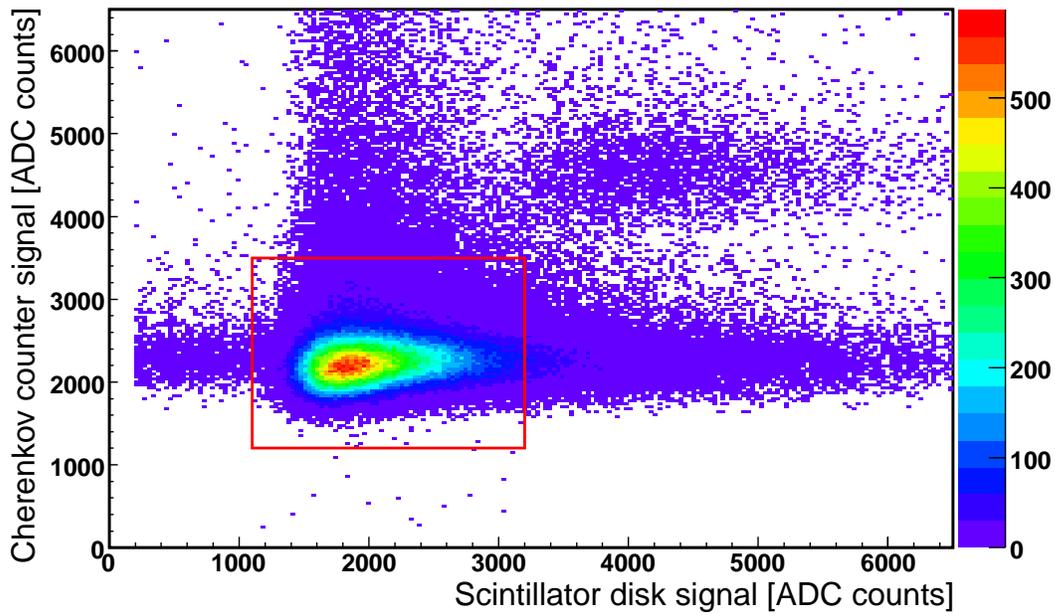}
  \caption{ Correlation between the signals of the Cherenkov counter and the scintillator disk counter. Events within the square were selected as a single beam particles.  }
\label{fig:chervsdisk}
\end{figure*}

A single particle candidate in the Cherenkov counter was validated by a time coincident detection in the scintillator disk counter at the entrance of the AIRFLY chamber. In addition, each candidate in a train was required to be well isolated in a window of $\pm 40$~ns, with no other single particle signal in the Cherenkov and scintillator disk counters. In the same time window, veto counters were required to have less than 10\% of their single particle signal in order to reject off-axis beam particles.   
Lastly, the whole train was rejected if more than four single particle candidates were found in the Cherenkov counter and/or in the scintillator disk counter.  Based on these strict criteria, about 25\% of all trains were selected, with a multiplicity of about 1.1 beam particle per train. To illustrate the quality of the single particle selection, the correlation between the signals of the Cherenkov and the scintillator disk counters is shown in Fig.~\ref{fig:chervsdisk} for one run. Events falling in the square are selected as single beam particles. Events accumulating in the upper right part of the figure, well separated from the selection area, correspond to two beam particles passing in the counters, and amount to only $\approx1\%$ of all events.
Events in the band with scintillator disk signal less than $\approx1000$~ADC counts, consistent with a single beam particle in the Cherenkov counter, are produced by particles hitting the less efficient border of the scintillator disk, and amount to only $\approx 0.5\%$ of all events. From these data, the contamination in the selected sample from events of multiplicity higher than one is estimated to be negligible. 
 
\begin{figure*}[!h]
\centering \includegraphics[width=5.8in]{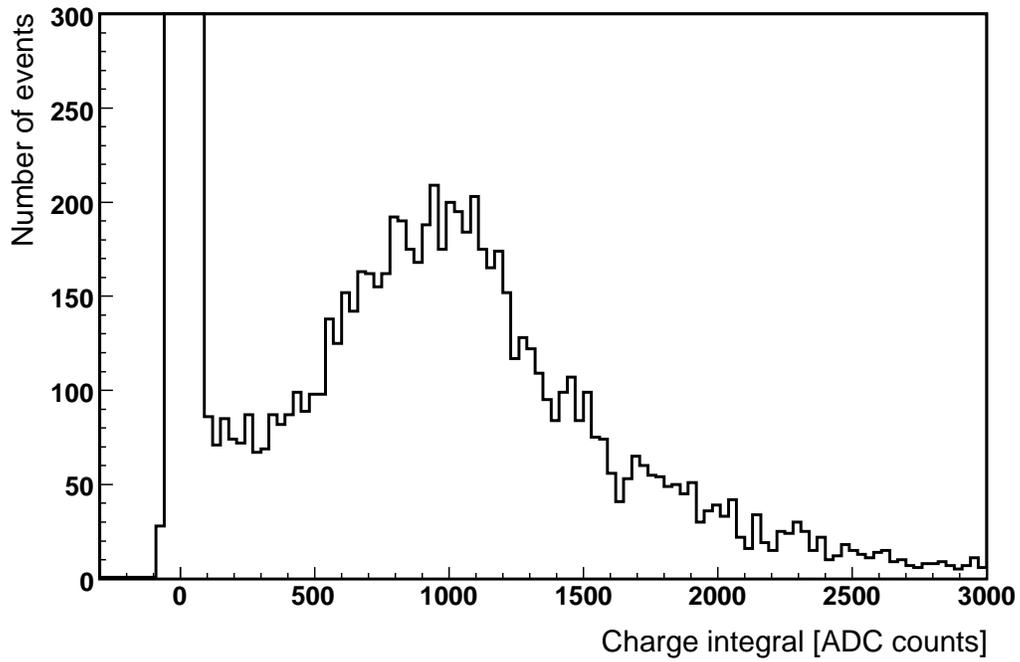}
  \caption{The distribution of the ADC charge integral corresponding to peaks in the PMT trace. A single photoelectron signal is evident.}
\label{fig:singlepe}
\end{figure*}

Then, a single photoelectron in time coincidence with the beam particle was searched for in the PMT trace. The distribution of the ADC charge integral corresponding to peaks in the PMT trace is shown in Fig.~\ref{fig:singlepe} for one data run taken in the Cherenkov configuration. To count a single photoelectron, the charge integral was required to be between 300 and 3000~ADC counts.  

\newpage
\newpage

\begin{figure*}[!t]
 \centering \includegraphics[width=5.675in]{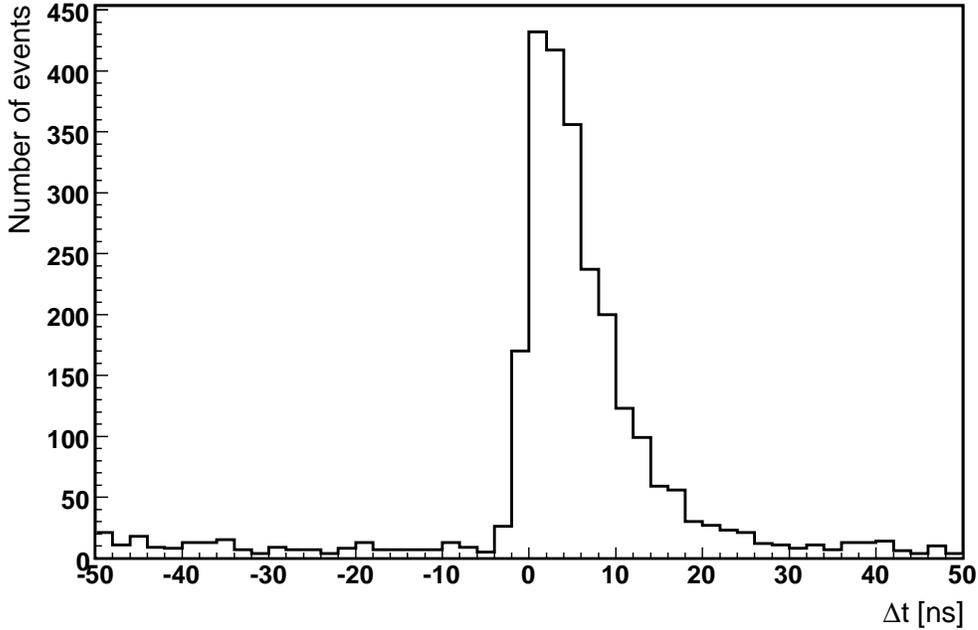}
  \caption{Distribution of the arrival times of single photoelectrons relative to the arrival time of the beam particle. }
\label{fig:singlepetime}
\end{figure*}

The arrival time of the selected photoelectrons relative to the beam particle, $\Delta\rm{t}$, is shown in  Fig. \ref{fig:singlepetime}. A clear signal is observed around $\Delta\rm{t}=0$ extending up to 30~ns. The exponential tail in the time distribution is due to the integrating sphere, since photons experience multiple diffusions before detection. Notice that a small background of single photoelectrons from accidentals is present before and after the main peak. The total number of single photoelectrons in a data taking run was calculated by integrating the time distribution in the interval [-4~ns, 32~ns], and subtracting the accidentals background as estimated from counts before -4~ns.

The rate of single photoelectrons, in units of \pbp, was then obtained by dividing the total number of single photoelectrons by the total number of single beam particles selected in the run. Results from several runs taken in the same configuration were averaged to improve the statistical uncertainty.  In the following, the rates measured in a given gas will be denoted by $R^{\rm{gas}}_{\rm{Fl}}$ and $R^{\rm{gas}}_{\rm{Ch}}$ for the Fluorescence and Cherenkov configurations, respectively.
 
Results are summarized in Tables \ref{tab:ratediffuser} to \ref{tab:ratelaser}, where the quoted uncertainties are statistical only. 
The rates of single photoelectrons measured with the diffuser plugs are reported in Table \ref{tab:ratediffuser}. Also, measurements with the chamber under vacuum yielded: 
\begin{eqnarray}
R_{\rm{Fl}}^{\rm{vac}} = (0.48 \pm 0.09) \cdot 10^{-4} ~\pbp,  \label{eq:vacfl}\\
R_{\rm{Ch}}^{\rm{vac}} = (3.11 \pm 0.21) \cdot 10^{-4} ~\pbp.
\label{eq:vacch}
\end{eqnarray}

The rates of single photoelectrons measured with the mirror plugs are reported in Table \ref{tab:ratemirror}.
 \begin{table}[ht] 
 \centering \begin{tabular}{|c|c|c|c|c|c|} \hline 
   $R^{\rm{N_2}}_{\rm{Fl}}$    &  $R^{\rm{air}}_{\rm{Fl}}$    &  $R^{\rm{He}}_{\rm{Fl}}$     &    $R^{\rm{N_2}}_{\rm{Ch}}$     &  $R^{\rm{air}}_{\rm{Ch}}$    &  $R^{\rm{He}}_{\rm{Ch}}$    \\ [0.5ex]	%
      ($10^{-4}~\pbp$)       &    ($10^{-4}~\pbp$)  &  ($10^{-4}~\pbp$)  &   ($10^{-4}~\pbp$)  &   ($10^{-4}~\pbp$)  &   ($10^{-4}~\pbp$)  \\ 	%
\hline 
$20.05 \pm 0.11$ & $3.22 \pm 0.07$ & $ 1.11 \pm 0.06$ & $32.89 \pm 0.15$ & $16.07 \pm 0.21$ & $3.78 \pm 0.11$   \\
 \hline 
\end{tabular}
\caption{Summary of the rates of single photoelectrons measured with the diffuser plugs.  
}
\label{tab:ratediffuser} 
 \end{table}
\begin{table}[ht] 
 \centering \begin{tabular}{|c|c|c|c|} \hline 
   $R^{\rm{N_2}}_{\rm{Fl}}$     &  $R^{\rm{He}}_{\rm{Fl}}$     &    $R^{\rm{N_2}}_{\rm{Ch}}$   &  $R^{\rm{He}}_{\rm{Ch}}$    \\ [0.5ex]	%
   ($10^{-4}~\pbp$)       &    ($10^{-4}~\pbp$)  &  ($10^{-4}~\pbp$)  &   ($10^{-4}~\pbp$)  \\ 	%
\hline 
$17.66 \pm 0.19$ & $1.19 \pm 0.09$ & $ 25.78 \pm 0.21$ & $1.05 \pm 0.09$   \\
 \hline 
\end{tabular}
\caption{Summary of the rates of single photoelectrons measured with the mirror plugs.
}
\label{tab:ratemirror} 
 \end{table}
A set of runs was dedicated to fluorescence measurements with an associated UV laser calibration (Section \ref{sec:lasersystem}). The diffuser plugs were used for these measurements. Results are summarized in Table \ref{tab:ratelaser}. For the measurement of the laser calibration constant, $S_{\rm{laser}} $, single photoelectrons, selected with the same criteria as the fluorescence measurement, were counted in time coincidence with the laser shots.
\begin{table}[ht] 
 \centering \begin{tabular}{|c|c|c|} \hline 
   $R^{\rm{N_2}}_{\rm{Fl}}$    &  $R^{\rm{He}}_{\rm{Fl}}$   &   $S_{\rm{laser}} $ \\ 	%
   ($10^{-4}~\pbp$)       &    ($10^{-4}~\pbp$)  &   ($10^{-4}~\pp$)  \\ 	%
\hline 
$19.31 \pm 0.14$ & $1.20 \pm 0.08$ & $ 10.70 \pm 0.11$   \\
 \hline 
\end{tabular}
\caption{Summary of measured fluorescence rates  with an associated UV laser calibration. $S_{\rm{laser}} $ is the measured laser calibration constant.}
\label{tab:ratelaser} 
 \end{table}

\section{Fluorescence yield measurement method}
\label{sec:abscherana}
The first step in the measurement of the fluorescence yield is the determination of the fluorescence rate in the 337~nm band, $S^{gas}_{\rm{Fl}}$,  for the atmospheric gas under study. The rate, $R_{\rm{Fl}}^{gas}$, measured in the Fluorescence configuration is given by:
 \begin{equation}
R_{\rm{Fl}}^{gas} = S_{\rm{Fl}}^{gas}+ B_{\rm{Fl}}^{gas},
\label{eq:fluor1}
\end{equation}
where B$^{gas}_{\rm{Fl}}$ accounts for the background. Sources of background could include particles - from the beam halo or from secondary interactions - producing Cherenkov and scintillation light in the PMT glass, in the interference filter and in the PTFE diffuser, or producing fluorescence in the small air volume between the PMT and the filter.
The background can be subtracted with high accuracy  by combining measurements in nitrogen and air.  
In fact, since the production of secondary particles  - mainly electrons - is practically the same in nitrogen and air, and particle interactions in the apparatus and its surroundings do not depend on the gas filling the chamber,  $B^{\rm{N_2}}_{\rm{Fl}} = B^{\rm{air}}_{\rm{Fl}}=  B_{\rm{Fl}}$ holds. Thus, backgrounds cancel in the difference, $\Delta R_{\rm{Fl}} $, between the rates measured in nitrogen and air:
 \begin{equation}
\Delta R_{\rm{Fl}} =  R_{\rm{Fl}}^{\rm{N_2}} - R_{\rm{Fl}}^{\rm{air}} =  S^{\rm{N_2}}_{\rm{Fl}} - S^{\rm{air}}_{\rm{Fl}} = S^{\rm{N_2}}_{\rm{Fl}} \left( 1 - \frac{1}{r_{\rm{N_2}}} \right),
\label{eq:fluor3}
\end{equation}
with $r_{\rm{N_2}}$ given by Eq.~\ref{eq:rn2}. The fluorescence rate in nitrogen, $S^{\rm{N_2}}_{\rm{Fl}}$, or equivalently the rate in air, ${S^{\rm{air}}_{\rm{Fl}} = S^{\rm{N_2}}_{\rm{Fl}}/r_{\rm{N_2}}}$, is obtained from  Eq. \ref{eq:fluor3} and the measured $\Delta R_{\rm{Fl}}$.

The measured fluorescence rate is converted into a fluorescence yield by two independent calibration methods. The first method is based on the calibrated 337~nm light of a nitrogen laser. The procedure to derive the laser calibration constant, $S_{\rm{laser}}$, is described in detail in Section \ref{sec:lasersystem}.  The second method is based on the known amount of Cherenkov light produced by the beam particle. 
Both Cherenkov and fluorescence emission contribute in the Cherenkov configuration, and the measured rate, $R_{\rm{Ch}}^{gas}$ is given by: 
 \begin{equation}
R_{\rm{Ch}}^{gas} = S^{gas}_{\rm{Ch}}+ B^{gas}_{\rm{Ch}} + S^{gas}_{\rm{Fl}}+B^{gas}_{\rm{Fl}},
\label{eq:cher1}
\end{equation}
where $S^{gas}_{\rm{Ch}}$ is the rate due to Cherenkov light and 
$B^{gas}_{\rm{Ch}}$ accounts for background originating from interactions of the beam particles in the exit port plug.
To separate the Cherenkov contribution, the difference, $\Delta R_{\rm{Ch-Fl}}^{gas}$, between the rates measured in the same gas with the Cherenkov and the Fluorescence configuration (Eqs. \ref{eq:fluor1} and \ref{eq:cher1}) is taken:
 \begin{equation}
\Delta R_{\rm{Ch-Fl}}^{gas} = R^{gas}_{\rm{Ch}}-R^{gas}_{\rm{Fl}}= S^{gas}_{\rm{Ch}}+B^{gas}_{\rm{Ch}}.
\label{eq:cherbkg1}
\end{equation}
When diffuser plugs are used, $B^{gas}_{\rm{Ch}}$ is dominated by Cherenkov light produced in the exit port plug and diffused back into the integrating sphere. When mirror plugs are used, a much smaller background from transition radiation by the thin aluminized mylar foil  is expected. In both cases, the background does not depend on the particular gas filling the chamber. Secondary particles, produced by the interaction of the beam with the exit port plug,  are another potential source of background, since they could induce emission of fluorescence light in the gas and of Cherenkov light in the sphere's diffuser. This background contribution, which would depend on the gas, was found to be negligible from both Monte Carlo simulations and cross-checks with the data (Section \ref{sec:systcher}). Thus,  $B^{gas}_{\rm{Ch}} = B_{\rm{Ch}}$ holds, and  the background can be effectively eliminated by an appropriate combination of measurements with different gases. In particular, measurements with helium,  which  accompanied every measurement with atmospheric gases, are used to measure the difference: 
 \begin{equation}
\Delta R_{\rm{Ch-Fl}}^{\rm{N_2}} - \Delta R_{\rm{Ch-Fl}}^{\rm{He}}=  S^{\rm{N_2}}_{\rm{Ch}} - S^{\rm{He}}_{\rm{Ch}}. \label{eq:chersig1}
\end{equation}
The Cherenkov rate in nitrogen is derived from Eq. \ref{eq:chersig1}, accounting for the small Cherenkov signal in helium ($S^{He}_{\rm{Ch}}/ S^{\rm{N_2}}_{\rm{Ch}} = 1\%$).

The fluorescence rate,   $S^{\rm{air}}_{\rm{Fl}}$, and the calibration rate, $S_{\rm{cal}}= S^{\rm{N_2}}_{\rm{Ch}} $ or $S_{\rm{laser}}$, are then combined in the ratio  $S^{\rm{air}}_{\rm{Fl}}/ S_{\rm{cal}} $. Notice that the absolute photon detection efficiency of the experiment cancels in the ratio.  Thus, systematic uncertainties of the fluorescence yield associated with the efficiency of the integrating sphere, the transmission of the interference filter, and most notably the PMT quantum efficiency are significantly reduced. 
 An equivalent ratio  of $ S^{\rm{MC}}_{\rm{Fl}}$ to $S^{\rm{MC}}_{\rm{cal}}=  S^{\rm{MC}}_{\rm{Ch}}$ or $ S^{\rm{MC}}_{\rm{laser}}$ is obtained from the Monte Carlo simulation of the experiment (Section \ref{subsec:simresults}) using a nominal fluorescence yield  $Y^{\rm{MC}}_{337} =   5.15  ~\rm{photons/MeV}$. 

The fluorescence yield of the 337~nm band in air is then measured from:
 \begin{equation}
 Y_{337}^{\rm{cal} }  =  \frac{  (S^{\rm{N_2}}_{\rm{Fl}}/  r_{\rm{N_2}})  /  S_{\rm{cal}} }{ S^{\rm{MC}}_{\rm{Fl}} /  S^{\rm{MC}}_{\rm{cal}} } Y^{\rm{MC}}_{337} .
 \label{eq:yieldth}
\end{equation}

\section{Fluorescence yield with Cherenkov calibration}
\label{sec:abscher}
In this Section, measurements of the fluorescence yield  based on  Cherenkov light for absolute calibration are presented. Two measurements were performed, with a slightly different apparatus (Section \ref{sec:sphere}). In the first, Cherenkov light emitted by the beam particle was diffused back into the integrating sphere by the diffuser mounted on the exit port plug. In the second measurement, the Cherenkov light was reflected  back into the sphere by the aluminized mylar mirror stretched over the exit port plug. In the following, details of the measurements are given. 

\subsection{Measurement with the diffuser plugs}
\label{sec:abschergore}
The analysis procedure detailed in Section \ref{sec:abscherana} was used to measure the fluorescence yield,  using data collected with the diffuser plugs mounted on the top and exit ports of the integrating sphere  (Table \ref{tab:ratediffuser}).

The fluorescence rate in nitrogen is derived from Eq. \ref{eq:fluor3} and the measured difference:  
\begin{equation}
\Delta R_{\rm{Fl}}= R_{\rm{Fl}}^{\rm{N_2}} - R_{\rm{Fl}}^{\rm{air}} = (16.83 \pm 0.13)\cdot 10^{-4}~\pbp, 
\label{eq:deltaFL}
\end{equation}
giving:
\begin{equation}
 S^{\rm{N_2}}_{\rm{Fl}} = (19.44 \pm 0.15)\cdot 10^{-4} ~\pbp.
\label{eq:fluorsignair}
\end{equation}
 Also, the fluorescence background is estimated from Eq.~\ref{eq:fluor1} as:
\begin{equation}
 B_{\rm{Fl}} = (0.61 \pm 0.07)\cdot 10^{-4} ~\pbp,
\label{eq:fluorbkg}
\end{equation}
which is only 3\% of the measured rate $R_{\rm{Fl}}^{\rm{N_2}}$.

The Cherenkov rate is derived from the difference (Eq.~\ref{eq:chersig1}): 
\begin{equation}
\Delta R_{\rm{Ch-Fl}}^{\rm{N_2}} - \Delta R_{\rm{Ch-Fl}}^{\rm{He}}=  S^{\rm{N_2}}_{\rm{Ch}} - S^{\rm{He}}_{\rm{Ch}} = (10.17 \pm 0.23) \cdot 10^{-4} ~\pbp.\label{eq:chersigm}
\end{equation}
Notice that  $R_{\rm{Fl}}^{\rm{He}}=  (1.11 \pm 0.06) \cdot 10^{-4} ~\pbp $ is larger than the fluorescence background of Eq. \ref{eq:fluorbkg},  indicating a nonzero $S^{\rm{He}}_{\rm{Fl}} $. Helium measurements were found to be very stable over the beam data taking period, suggesting a residual nitrogen contamination in the helium gas bottle\footnote{Nitrogen impurities in helium are known to produce sizable fluorescence emission~\cite{nitrogenhelium}. Also, we observed a strong signal when introducing in our chamber a few hPa of nitrogen gas in helium at atmospheric pressure.}.  This small fluorescence-like signal in helium does not affect  Eq.~\ref{eq:chersigm}, since   $S^{\rm{He}}_{\rm{Fl}} $ cancels in $\Delta R_{\rm{Ch-Fl}}^{\rm{He}}$.
 After correcting Eq. \ref{eq:chersigm} for the small Cherenkov signal in helium, the Cherenkov rate in nitrogen is obtained:
\begin{equation}
 S^{\rm{N_2}}_{\rm{Ch}} = (10.27 \pm 0.23) \cdot 10^{-4} ~\pbp \label{eq:chersign2}.
\end{equation}
The Cherenkov background is estimated as: 
\begin{equation}
 B_{\rm{Ch}} = \Delta R_{\rm{Ch-Fl}}^{\rm{He}}  - S^{\rm{He}}_{\rm{Ch}} = (2.57 \pm 0.13) \cdot 10^{-4} ~\pbp.\label{eq:cherbkg}
\end{equation}
An independent measurement of the Cherenkov signal was obtained from runs performed in the Cherenkov configuration with air filling the chamber. Following the same procedure used to derive the signal in nitrogen,  the Cherenkov rate in air was measured to be:
\begin{equation}
 S^{\rm{air}}_{\rm{Ch}} = (10.28 \pm 0.26) \cdot 10^{-4}~ \pbp.\label{eq:chersigair}
\end{equation}
Due to the slightly different index of refraction,  the Cherenkov yield in nitrogen is 1.7\% greater than the one in air, well within the uncertainty of the measurements.
In the Monte Carlo simulation, the pressure and temperature of the air volume were chosen to match the average index of refraction of the air and nitrogen measurements. Thus, the best estimate of the Cherenkov rate is given by the average of Eq.~\ref{eq:chersign2} and Eq.~\ref{eq:chersigair}:
\begin{equation}
 S^{\rm{avg}}_{\rm{Ch}} = (10.27 \pm 0.19) \cdot 10^{-4} ~\pbp \label{eq:chersigavg},
\end{equation}
in very good agreement with the Monte Carlo simulation ($S^{\rm{MC}}_{\rm{Ch}}=10.53 \cdot 10^{-4}~\pbp$, Table  \ref{tab:rateMC}). 
This agreement, while not essential for the fluorescence yield measurement since the absolute photon detection efficiency cancels in the ratio of the fluorescence to Cherenkov rate, confirms the reliability of the simulation in reproducing the overall efficiency of the experiment.

 Equation~\ref{eq:yieldth}, with ${S_{\rm{cal}}=S^{\rm{avg}}_{\rm{Ch}}}$ and  ${S^{\rm{MC}}_{\rm{cal}}=S^{\rm{MC}}_{\rm{Ch}}}$, is then used to derive the fluorescence yield. From the measured fluorescence and Cherenkov rates (Eqs. \ref{eq:fluorsignair} and \ref{eq:chersigavg}),  and the corresponding Monte Carlo rates of Table \ref{tab:rateMC}, the fluorescence yield of the 337~nm band in air is found to be:
 \begin{equation}
 Y_{337}^{\rm{ChDiff} }  =  5.64 \pm 0.12   ~\rm{photons/MeV.}\label{eq:flres1}
\end{equation}

\subsection{Measurement with the mirror plugs}
\label{sec:abschermyl}
Data collected with the mirror plugs mounted on the top and exit ports of the integrating sphere (Table \ref{tab:ratemirror}) were also analyzed following the procedure described in Section \ref{sec:abscherana}. 

The fluorescence rate in nitrogen is obtained by subtracting the background of  Eq.~\ref{eq:fluorbkg}  from the measured signal in the Fluorescence configuration: 
\begin{equation}
 S^{\rm{N_2}}_{\rm{Fl}} = R^{\rm{N_2}}_{\rm{Fl}} - B_{\rm{Fl}} = (17.05 \pm 0.21)\cdot 10^{-4} ~\pbp.
\label{eq:fluorsignairmylar}
\end{equation}

The Cherenkov rate is derived from the difference (Eq.~\ref{eq:chersig1}): 
\begin{equation}
\Delta R_{\rm{Ch-Fl}}^{\rm{N_2}} - \Delta R_{\rm{Ch-Fl}}^{\rm{He}}=  S^{\rm{N_2}}_{\rm{Ch}} - S^{\rm{He}}_{\rm{Ch}} = (8.26 \pm 0.31) \cdot 10^{-4} ~\pbp.\label{eq:chersigmylar1}
\end{equation}
After correcting Eq.~\ref{eq:chersigmylar1} for the small Cherenkov signal in helium, the Cherenkov rate in nitrogen is obtained:
\begin{equation}
 S^{\rm{N_2}}_{\rm{Ch}} = (8.34 \pm 0.31) \cdot 10^{-4} ~\pbp \label{eq:chersignmylar2},
\end{equation}
in very good agreement with the Monte Carlo simulation ($S^{\rm{MC}}_{\rm{Ch}}=8.27 \cdot 10^{-4}~\pbp$, Table  \ref{tab:rateMC}). This agreement gives further confidence on the quality of the simulation.

The fluorescence yield is then derived from Eq.~\ref{eq:yieldth}, with $S_{\rm{cal}}=S^{\rm{N_2}}_{\rm{Ch}}$ and  $S^{\rm{MC}}_{\rm{cal}}=S^{\rm{MC}}_{\rm{Ch}}$.  From the measured fluorescence and Cherenkov rates (Eqs. \ref{eq:fluorsignairmylar} and \ref{eq:chersignmylar2}), and the corresponding Monte Carlo rates of Table \ref{tab:rateMC}, the fluorescence yield of the 337~nm band in air is found to be:
 \begin{equation}
 Y_{337}^{\rm{ChMirr} } =   5.48 \pm 0.25   ~\rm{photons/MeV,}\label{eq:flres2}
\end{equation}
 in very good agreement with $ Y_{337}^{\rm{ChDiff} } $ (Eq. \ref{eq:flres1}). 
 
Notice that the absolute photon detection efficiency is $\approx 20\%$ smaller with the mirror plugs (Table \ref{tab:rateMC}). Also, there is a significant difference between the Cherenkov background with the mirror plugs, estimated as $\Delta R_{\rm{Ch-Fl}}^{\rm{He}}  = (-0.14 \pm 0.13) \cdot 10^{-4} ~\pbp$, and that with the diffuser plugs,  $ B_{\rm{Ch}} = 2.57 \cdot 10^{-4} ~\pbp$ (Eq. \ref{eq:cherbkg}). The agreement between 
$Y_{337}^{\rm{ChMirr} } $  and $Y_{337}^{\rm{ChDiff} } $ confirms the reliability of the Cherenkov calibration method. 

\subsection{Cross-checks and systematic uncertainties}
\label{sec:systcher}
Systematic uncertainties  are largely correlated in the two Cherenkov calibration measurements,
which use the same apparatus - except for the top and exit port plugs of the integrating sphere - and data analysis technique. In the following, we will refer to $ Y_{337}^{\rm{Ch}} $ when cross-checks and systematic uncertainties apply to both $ Y_{337}^{\rm{ChDiff} }$ and $ Y_{337}^{\rm{ChMirr} }$.

Fluorescence of the diffuser material could potentially bias the fluorescence yield measurement. Negligible fluorescence induced by UV light, advertised by the GORE$\texttrademark$ DRP$^{\textregistered}$ manufacturer, was confirmed by our measurements with a deuterium lamp. In addition, fluorescence produced by charged particles traversing the diffuser was also measured with an 18~$\mu$Ci~$^{241}$Am source placed at the entrance port of the sphere. The chamber was evacuated, letting alpha particles hit the inner surface of the sphere. In these conditions, the PMT single photoelectron rate,  $\approx 800 $~Hz, was found to be independent of the presence of the $^{241}$Am source. The PMT rate increased to 20~kHz when nitrogen at atmospheric pressure filled the chamber, due to fluorescence emission from the nitrogen molecules excited by the alpha particles. From these results, fluorescence of the diffuser induced by charged particles, if any, can be safely considered negligible as far as the $ Y_{337}^{\rm{Ch}} $  measurement is concerned. 

In the Fluorescence configuration, Cherenkov light produced by the beam particles is dumped on the black UV absorbing material covering the inner surface of the chamber. A fraction of this light could get back into the sphere through its open exit port, introducing a bias on $ Y_{337}^{\rm{Ch}} $. To investigate this potential systematic effect, a 337~nm laser beam was shot along the beam line into the Cherenkov dump, and the corresponding light intensity at the PMT port was measured. The light intensity was found to be $<10^{-3}$  of that measured when the laser hit the sphere's diffusive wall. Since the nitrogen fluorescence light collected by the sphere is about twice the Cherenkov light (Sections \ref{sec:abschergore} and \ref{sec:abschermyl}), the effect of Cherenkov light bouncing back into the sphere from the Cherenkov light dump is negligible.

Several checks of the data analysis and background subtraction method were performed.    $ Y_{337}^{\rm{Ch}}$ was found not to depend on the value of the threshold chosen to select single photoelectrons. This result was expected, since the same threshold is used in both fluorescence and Cherenkov measurements.     
A potential bias due to the slightly different time evolution of the Cherenkov and fluorescence signals was studied by changing the time window for photon counting between 24 and 40~ns, which produced differences in  $ Y_{337}^{\rm{Ch}}$ of less than 1\%. 
Much stricter selection criteria for isolated particles, which reduced the selected sample by a factor of two, resulted in a 1\% shift of   $ Y_{337}^{\rm{Ch}}$.  An independently developed data selection analysis, with a different peak finding algorithm and background rejection criteria, was also performed, which changed $ Y_{337}^{\rm{Ch}}$ by 0.7\%. Potential biases in the determination of the fluorescence and Cherenkov backgrounds were also investigated. For $  Y_{337}^{\rm{ChDiff} } $, measurements with the chamber evacuated provide an independent estimate of $B_{\rm{Fl}} $ and $B_{\rm{Ch}} $.  $R_{\rm{Fl}}^{\rm{vac}}=(0.48 \pm 0.09) \cdot 10^{-4}~\pbp$ (Eq. \ref{eq:vacfl}) is consistent within uncertainties with the fluorescence background of Eq.  \ref{eq:fluorbkg}, confirming the Monte Carlo expectation that background from secondaries produced in the chamber gas is negligible. An estimate of  $B_{\rm{Ch}} $ is derived from $ \Delta R_{\rm{Ch-Fl}}^{\rm{vac}}= (2.63 \pm 0.23) \cdot 10^{-4}~\pbp$, which is  consistent with Eq. \ref{eq:cherbkg}. Another check of  $B_{\rm{Ch}} $ is provided by $R^{\rm{N_2}}_{\rm{Ch}}-R^{\rm{air}}_{\rm{Ch}} =  (16.82 \pm 0.26) \cdot 10^{-4}~\pbp  $,  where the Cherenkov background in nitrogen and air should cancel in the difference. The result is consistent with  Eq.~\ref{eq:deltaFL} within their uncertainties, indicating that  Cherenkov background does not depend on the chamber gas, and fluorescence produced by secondaries from the exit port plug is negligible, as expected from Monte Carlo simulations.  For $  Y_{337}^{\rm{ChDiff} } $, the same fluorescence background (Eq.~\ref{eq:fluorbkg}) of the diffuser plugs measurement is used. This choice is justified by the stability of the helium measurements (Tables \ref{tab:ratediffuser} and \ref{tab:ratemirror}), performed routinely to monitor the background level. If the observed difference ($0.08 \pm  0.11~\pbp$) between the two $R^{\rm{He}}_{\rm{Fl}} $ rates is due to a change of the fluorescence background level,  the effect on $ Y_{337}^{\rm{ChDiff} }$ would only be of 0.5\%. 
All performed checks resulted in shifts of  $ Y_{337}^{\rm{Ch}}$ smaller than its statistical uncertainty, and  a 1\% systematic uncertainty is  conservatively assigned to the data selection and background subtraction methods. 

Recent studies \cite{ulrich} indicate that  the absolute value of $r_{\rm{N_2}}$ may depend on the presence of very small amounts of contaminants in the nitrogen gas. Thus, a bias in the fluorescence yield could be introduced by changes in the nitrogen gas composition during the measurements. To study the stability of $r_{\rm{N_2}}$, we performed measurements with different gas bottles, with different conditions of gas flow in the chamber and for extended periods of time. The observed variations are included in the 1\% uncertainty quoted in Eq.~\ref{eq:rn2}. Also, the rates $R^{\rm{N_2}}_{\rm{Fl}}$ measured during several weeks at the Test Beam did not show any noticeable change beyond statistical fluctuations. From these studies, a 1\% systematic uncertainty is assigned to $ Y_{337}^{\rm{Ch}}$ for the nitrogen to air fluorescence ratio.

The absolute value of the photon detection efficiency, including the peak value of the integrating sphere's efficiency, of the 337 nm filter transmission and of the PMT quantum and collection efficiency, cancels in the ratio of fluorescence to Cherenkov signal, and its contribution to the systematic uncertainty of $ Y_{337}^{\rm{Ch}}$ is negligible.

A difference in the sphere's efficiency between the Fluorescence and Cherenkov configurations may introduce a bias.  This was studied by shooting a 337 nm laser beam on different points of the sphere's inner surface, and switching between the  Fluorescence and Cherenkov configurations. The maximum shift in the light intensity measured at the side port was 0.9\%, which is taken as an estimate of the systematic uncertainty on $ Y_{337}^{\rm{Ch}}$ associated to the sphere's efficiency.

The uncertainty associated to the wavelength dependence of the detection efficiency must also be evaluated, due to the wide wavelength range of the Cherenkov emission.  
In the Monte Carlo simulation of the experiment, the wavelength dependence of the diffuser is included in each diffusion bounce inside the sphere (Section \ref{subsec:photondet}). We also measured the overall efficiency of the integrating sphere between 300 and 400~nm by detecting at the side port the light spectrum of a deuterium lamp with an Oriel MS257$^{TM}$ spectrograph. This efficiency was then used in a simulation to weight photons exiting the side port. The two simulation approaches yielded a 1\% shift of $ Y_{337}^{\rm{Ch}}$, which is taken as an estimate of the systematic uncertainty.
The wavelength dependence of the PMT quantum efficiency was varied according to its uncertainty, and a corresponding 1\% systematic uncertainty is assigned. The  wavelength and angular dependence transmission of the 337 nm interference filter was measured by two independent groups; when used in the Monte Carlo simulation, the difference between the two measurements resulted in a 2\% shift  of $Y_{337}^{\rm{Ch}}$, which is taken as systematic uncertainty. The uncertainty in the relative intensities of the fluorescence bands \cite{airflyp} has a negligible effect, since the signal is dominated by the 337~nm band (Fig. \ref{fig:effdet}). 

From the studies of Section \ref{subsec:edepsim}, a 2\% systematic uncertainty is assigned to the simulation of the energy deposit, while the uncertainty of the Cherenkov yield from the gas index of refraction is considered negligible.  
The limited statistics of the Monte Carlo simulations contribute 1\% to the uncertainty.  

The different contributions to the systematic uncertainty are summarized in Table \ref{tab:syst_cere}. Taking these contributions as uncorrelated, a total systematic uncertainty of 3.7\% on $  Y_{337}^{\rm{Ch}} $ is obtained. Notice that all contributions should be considered fully correlated between $Y_{337}^{\rm{ChDiff} } $ and $ Y_{337}^{\rm{ChMirr} }$, with the exception of the Monte Carlo statistics since separate simulations were performed for the two setups. 

\begin{table}[ht]
 \begin{center}
 \begin{tabular}{|c|c|}
 \hline
data selection and  background subtraction            & 1.0\% \\
 $r_{\rm{N_2}}$          & 1.0\% \\
 integrating sphere efficiency        & 0.9\% \\
integrating  sphere wavelength dependence & 1.0\% \\
 PMT quantum efficiency      & 1.0\% \\
 filter transmittance        & 2.0\% \\
simulation of energy deposit                    & 2.0\% \\
 Monte Carlo statistics      & 1.0\% \\
 \hline
 Total                       & 3.7\% \\
 \hline
 \end{tabular}
 \caption{Systematic uncertainties on the fluorescence yield measured with Cherenkov calibration, $ Y_{337}^{\rm{ChDiff} }$ and $ Y_{337}^{\rm{ChMirr} }$. }\label{tab:syst_cere}
 \end{center}
 \end{table}

\newpage
\section{Fluorescence yield with laser calibration}
\label{sec:abslaser}
The fluorescence yield measurement reported in this Section is  largely independent of those based on Cherenkov calibration, since a nitrogen laser (Section~\ref{sec:lasersystem}) was used for the absolute calibration. 
Data presented in Table~\ref{tab:ratelaser} were analyzed with the procedure of Section~\ref{sec:abscherana}.  

The fluorescence rate in nitrogen is obtained by subtracting the background of  Eq.~\ref{eq:fluorbkg}  from the rate measured in the Fluorescence configuration:
\begin{equation}
 S^{\rm{N_2}}_{\rm{Fl}} = R^{\rm{N_2}}_{\rm{Fl}} - B_{\rm{Fl}} = (18.70 \pm 0.16)\cdot 10^{-4} ~\pbp.
\label{eq:fluorsignairlaser}
\end{equation} 

For the absolute calibration, the laser calibration constant (Table~\ref{tab:ratelaser}) 
 \begin{equation}
 S_{\rm{laser}}  = (10.70 \pm 0.11) \cdot   10^{-4}   ~\pp, \label{eq:laser}
\end{equation}
was obtained from  several calibration runs performed routinely at the end of each day of data taking.

The fluorescence yield is then derived from Eq. \ref{eq:yieldth}, with $S_{\rm{cal}}=S_{\rm{laser}}$ and  $S^{\rm{MC}}_{\rm{cal}}=S^{\rm{MC}}_{\rm{laser}}$.  From the measured fluorescence rate (Eq.~\ref{eq:fluorsignairlaser}), the laser calibration constant (Eq.~\ref{eq:laser}),  and the corresponding Monte Carlo rates of Table~\ref{tab:rateMC}, the fluorescence yield of the 337~nm band in air is found to be:
\begin{equation}
Y_{337}^{\rm{laser} }= 5.73 \pm 0.08   ~\rm{photons/MeV,} \label{eq:flres3}
\end{equation}
in very good agreement with the Cherenkov calibrated measurements $ Y_{337}^{\rm{ChDiff} }$ (Eq.~ \ref{eq:flres1}) and $ Y_{337}^{\rm{ChMirr} }$ (Eq.~\ref{eq:flres2}).

Notice that the fluorescence rate (Eq.~\ref{eq:fluorsignairlaser}) is somewhat smaller than the rate measured in the Cherenkov calibration measurement (Eq.~ \ref{eq:fluorsignair}). In fact, after completing the measurement with laser calibration, the apparatus was disassembled, the integrating sphere was cleaned and new diffuser plugs were installed, which resulted in a slightly higher efficiency of the sphere in the subsequent measurement with Cherenkov calibration. Since the efficiency of the sphere cancels out in Eq.~\ref{eq:yieldth}, the fluorescence measurement is not affected and the agreement between $Y_{337}^{\rm{laser} }$ and the measurements with Cherenkov calibration confirms the reliability of the calibration methods. 

\subsection{Cross-checks and systematic uncertainties}
\label{sec:syslaser}
Several of the systematic uncertainties evaluated in Section \ref{sec:systcher} are common to $Y_{337}^{\rm{laser} }$, since the same apparatus and data analysis techniques were used. In particular, systematic uncertainties from the data selection and background subtraction, the nitrogen to air fluorescence ratio, the integrating sphere's efficiency, and the simulation of the energy deposit are fully correlated between $Y_{337}^{\rm{laser} }$ and $Y_{337}^{\rm{Ch} }$. On the other hand, systematic uncertainties associated to the wavelength dependence of the integrating sphere's efficiency, to the PMT quantum efficiency, and to the interference filter were found to be negligible, since the nitrogen laser calibration light, unlike Cherenkov light, reproduces exactly the 337 nm fluorescence band. 

Checks on the background subtraction specific to the laser calibration measurement included the stability of the helium measurements (Tables~\ref{tab:ratediffuser} and \ref{tab:ratelaser}), performed routinely to monitor the background level. 
If the observed difference ($0.09 \pm  0.10~\pbp$) between the two $R^{\rm{He}}_{\rm{Fl}} $ rates is attributed to a change of the fluorescence background level,  the effect on $ Y_{337}^{\rm{laser} }$ would only be of 0.5\%. 
The background contribution from accidentals in the determination of $S_{\rm{laser}}$ was negligible, and does not contribute to the systematic uncertainty.

Systematic uncertainties specific to the laser calibration procedure included the knowledge of the geometry of the chamber. Since the fluorescence signal is proportional to the path length of the beam particle in the integrating sphere, $S^{\rm{MC}}_{\rm{Fl}}$ depends on the geometry introduced in the Monte Carlo simulation. Due to the mechanical assembly of the integrating sphere, which includes in its interior the soft PTFE layer of 0.5~mm thickness,  the uncertainty on the internal size of the sphere is conservatively taken as 0.5~mm, which corresponds to a 0.3\% systematic uncertainty on  $ Y_{337}^{\rm{laser} }$. Notice that this uncertainty is absent in the Cherenkov calibration measurements of Section \ref{sec:abscher}, since both fluorescence and Cherenkov light are proportional to the path length of the beam particle, which cancels in their ratio.  

 The absolute calibration of the laser probe and the efficiency of the laser calibration sphere introduce systematic uncertainties specifically associated with the laser calibration procedure. For  the absolute calibration of the RjP-465 probe, the 5\% uncertainty quoted by the manufacturer was taken. The reproducibility of the absolute calibration was checked by comparing two RjP-465 probes, which were found to agree within  2.6\%, well within the quoted systematic uncertainty. For the efficiency of the laser calibration sphere (Section \ref{sec:lasersystem}), the 0.5\% uncertainty associated to the linearity of the probe, combined with a 0.6\% uncertainty estimated from the reproducibility of the measurement, gave a 0.8\% systematic uncertainty on  $ Y_{337}^{\rm{laser} }$. 
 
 Other checks include the stability of $S_{\rm{laser}}$ and measurements with different particle beams. The laser calibration constant was measured shooting the laser beam on different spots in the integrating sphere, and repeating the measurement after disassembling and remounting the laser system. In all of these checks,  $S_{\rm{laser}}$ changed by less than 1\%. 
Measurements of the fluorescence yield with beams of 32~GeV pions and 8~GeV positrons were also performed during the test beam period dedicated to the laser calibration. The multiplicity of these secondary beams was significantly lower than the primary proton beam, and the corresponding fluorescence yield measurements are not statistically competitive. Nevertheless, they provide a useful cross-check of the independence of the fluorescence yield to the type of relativistic particle originating the energy deposit. Indeed, measurements performed with these secondary beams were consistent within their 6\% statistical uncertainty with the fluorescence yield of Eq.~\ref{eq:flres3}. 

 The different contributions to the systematic uncertainty are summarized in Table~\ref{tab:syst_laser}. Taking these contributions as uncorrelated, a total systematic uncertainty of 5.8\% on $  Y_{337}^{\rm{laser} } $ is obtained.    

 \begin{table}[htpb]
 \begin{center}
 \begin{tabular}{|c|c|}
 \hline
data selection and  background subtraction            & 1.0\% \\
 $r_{\rm{N_2}}$            & 1.0\% \\
integrating sphere efficiency        & 0.9\% \\
geometry                    & 0.3\% \\
laser probe calibration     & 5.0\% \\
calibration sphere transmission  & 0.8\% \\
simulation of energy deposit                    & 2.0\% \\
 Monte Carlo statistics      & 1.0\% \\
 \hline
 Total                       & 5.8\% \\
 \hline

 \end{tabular}
 \caption{Systematic uncertainties on the fluorescence yield measured with laser calibration, $Y_{337}^{\rm{laser} }$.}\label{tab:syst_laser}
 \end{center}
 \end{table}
 
\section{Combined fluorescence yield measurement}
\label{sec:combined}
The fluorescence yield measurements presented in Sections \ref{sec:abscher} and \ref{sec:abslaser} - $ Y_{337}^{\rm{ChDiff} }$, $ Y_{337}^{\rm{ChMirr} }$ and $ Y_{337}^{\rm{laser} }$ - are found to be in good agreement within their uncertainties. They are based on very different calibration light sources - photons from Cherenkov emission or from a nitrogen laser -  whose corresponding systematic uncertainties are largely uncorrelated. Thus, an appropriate weighted average of these measurements yields an improved result.

 First, the measured fluorescence yields were extrapolated to a nominal pressure  of 1013~hPa and temperature of 293~K. This procedure, which used our measured $p'_{air}(337)=15.89$~hPa \cite{airflyp} and  $\alpha_{337}=-0.36$ \cite{airflyth}, shifted the measurements by less than 0.5\%. Then, a weighted average was performed, taking into account the correlation between uncertainties. In particular, systematic uncertainties associated to the integrating sphere wavelength dependence, the PMT quantum efficiency, the filter transmittance, the geometry, the laser probe calibration, and the calibration sphere transmission were taken as uncorrelated between the Cherenkov and the laser calibration measurements. 
From the weighted average, the fluorescence yield of the 337~nm band in air at 1013~hPa and 293~K was found to be:
 \begin{equation}
 Y_{337}  = 5.61\pm 0.06_{stat} \pm 0.21_{syst}  ~\rm{photons/MeV,}\label{eq:flresfinal}
\end{equation}
where the statistical and systematic uncertainties are quoted separately. 

\begin{figure*}[th]
\centering \includegraphics[width=5.5in]{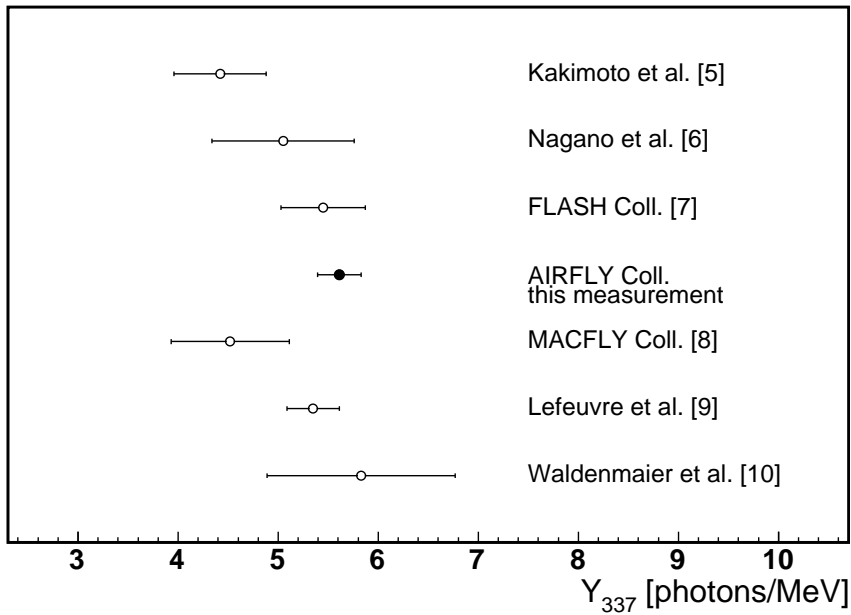}
  \caption{Experimental results on $Y_{337}$. For some experiments, the fluorescence yield of the 337~nm band is derived from the integrated yield measured between $\approx 300$ to 400~nm (see text for details).
}
\label{fig:fymeasurements}
\end{figure*}

Of the recent fluorescence yield experiments, only a few   \cite{nagano1} \cite{airlight} have measured the yield of the 337 nm band in air and can be directly compared with the AIRFLY measurement of Eq. \ref{eq:flresfinal}. Most of the experiments \cite{kakim} \cite{belz} \cite{colin} \cite{gorod} report a fluorescence yield integrated over the spectrum between $\approx 300$ to 400~nm. To compare with these experiments,  we convert the integrated yield into a yield of the 337~nm band by using the spectrum measured by AIRFLY~\cite{airflyp}. Results for $Y_{337}$ are compared in Fig.~\ref{fig:fymeasurements}. The AIRFLY measurement is compatible with previous measurements
and presents the smallest uncertainty. It is also compatible with the detailed study by \cite{arquerossyst}, where it is argued that some of these measurements should be corrected for a systematic bias in their calculation of the energy deposit.

\section{Conclusions}
\label{sec:conclusions}
We have performed a precise measurement of the absolute fluorescence yield of the 337 nm band in air relevant for UHECR experiments. 
The fluorescence emission was calibrated by a known light source - Cherenkov emission from the beam particle or a calibrated nitrogen laser - measured in the same apparatus. 
With this novel experimental method, the impact of the photomultiplier detection efficiency was minimized, reducing significantly the systematic uncertainty. Two independent calibrations provided consistent results, and a total uncertainty of 4\% on the absolute fluorescence yield was achieved. 

The AIRFLY measurements - the absolute yield reported here and the pressure, temperature and humidity dependence of the fluorescence spectrum~\cite{airflyp} \cite{airflyth} - provide the most comprehensive and precise parameterization of the fluorescence yield currently available. These measurements have direct implications for UHECR experiments which employ Fluorescence Detectors to determine the cosmic ray energy. For example, the absolute fluorescence yield of the 337 nm band reported here is 11\% and 30\% larger than that currently adopted by the Pierre Auger Observatory~\cite{augersyst} \cite{augerspectrum} and by the Telescope Array~\cite{tasyst} \cite{taspectrum}, respectively.
While the actual effect on the UHECR energy spectrum also depends on the specific fluorescence spectrum adopted by these experiments, a downward shift of the energy scale by at least $\approx 10\%$ is implied by the AIRFLY result.  At the same time, the uncertainty on the energy scale associated to the fluorescence yield, currently a major contribution~\cite{augersyst}~\cite{tasyst}, will be reduced by a factor of about three.

In principle, the experimental methods developed by AIRFLY could be further refined to improve the precision of the fluorescence yield. In particular, the 5\% systematic uncertainty of the laser energy probe - the main systematic of the pulsed laser calibration method - may be reduced, or a continuous laser absolutely calibrated to 1-2\% could be employed.  However,   the uncertainty on the energy scale of UHECR experiments is likely to be dominated by other contributions, including the absolute calibration of the fluorescence telescopes and the knowledge of the atmosphere. 
Thus, we expect the absolute fluorescence yield measured by AIRFLY to remain a reference for the current and next generation of UHECR experiments. 

\section{Acknowledgements}
We thank J. Appel, L. Bellantoni, D. Jensen, E. Ramberg, and A. Soha for their continuous support during the measurements at the FNAL Test Beam Facility. We acknowledge the stimulating discussions with the participants to the Air Fluorescence Workshops on theoretical and experimental aspects of the fluorescence yield measurement. We thank F. Arqueros and J. Rosado for performing independent checks of our simulations and for their careful reading of the manuscript.  This work was supported in part by the Kavli Institute for Cosmological Physics at the University of Chicago through grants NSF PHY-0114422 and NSF PHY-0551142 and an endowment from the Kavli Foundation and its founder Fred Kavli; by the University of Chicago and the Department of Energy under section H.44 of Department of Energy Contract No. DE-AC02-07CH11359 awarded to Fermi Research Alliance, LLC and Contract No. DE-AC02-06CH11357 awarded to UChicago Argonne, LLC, USA; by grant LA08016 of MSMT CR, Czech Republic; by the BMBF with Contract No. 05A08VK1, Germany; by the Istituto Nazionale di Fisica Nucleare (INFN), Italy.



\end{document}